\documentclass[12pt]{article}

\usepackage[margin=2.5cm]{geometry}

\usepackage{amsmath,amsthm,amsfonts,amscd,amssymb,bbm,mathrsfs,enumerate,url}
\usepackage{csquotes}

\usepackage{graphicx,tikz}
\usetikzlibrary{matrix,arrows}
\usetikzlibrary{decorations.pathreplacing}
\usetikzlibrary{decorations.pathmorphing}
\usetikzlibrary{patterns,fadings}

\def\R
R{{\mathbb R}}

\def\R{{\mathrm R}}

\def\1{{\mathbbm 1}}

\def\diff{{\rm Diff}}

\def\diffs1{\diff(S^1)}

\def\psl2r{{\rm PSL}(2,\RR)}
\def\sl2r{{\rm SL}(2,\RR)}
\def\su11{{\rm SU}(1,1)}
\def\2dmob{{\overline{\psl2r}\times\overline{\psl2r}}}
\def\<{\langle}
\def\>{\rangle}

\newcommand{\svh}{\mathrm{Sv/h}}

\newif\ifLetterStyle

\LetterStylefalse

\title{Comments on
``Individual external dose monitoring of all citizens of Date {C}ity by
  passive dosimeter 5 to 51 months after the {F}ukushima {NPP} accident (series):
  1. {C}omparison of individual dose with ambient dose rate monitored by aircraft
  surveys.''
}
\date{} 

\newcommand{\myauthors}{
{\bf Masaki Oshikawa} \\
Institute for Solid State Physics, University of Tokyo\\
email: {\tt oshikawa@issp.u-tokyo.ac.jp}\\
{\bf Yutaka Hamaoka} \\
Faculty of Business and Commerce, Keio University\\
email: {\tt hamaoka@fbc.keio.ac.jp}\\
{\bf Kyo Kageura} \\
Graduate School of Education, University of Tokyo\\
email: {\tt kyo@p.u-tokyo.ac.jp}\\
{\bf Shin-ichi Kurokawa} \\
The High Energy Accelerator Research Organization (KEK), Tsukuba \\
email: {\tt shin-ichi.kurokawa@kek.jp}\\
{\bf Jun Makino} \\
Department of Planetology, Graduate School of
  Science, Kobe University\\
email: {\tt makino@mail.jmlab.jp}\\
{\bf Yoh Tanimoto}
\\
   Dipartimento di Matematica, Universit\`a di Roma Tor Vergata\\
   email: {\tt hoyt@mat.uniroma2.it}
}

\ifLetterStyle
\author{}
\else 
\author{\myauthors}
\fi 

\begin{document}
\maketitle
%

\ifLetterStyle
\vspace{0.3cm}
\noindent
Dear Sir,

\vspace{0.2cm}
\fi

In this Letter, we point out inconsistencies, obvious mistakes and inappropriate statements
in the paper \cite{MH16} published in \textit{Journal of Radiological Protection} 
.
\begin{enumerate}[{(}1{)}]
 \item In Ethics Statement, the authors write ``\textit{[T]he geocoded household addresses of the glass-badge monitoring
 participants were pseudo-anonymized by rounding both longitude and latitude coordinates to $1/100$
 degrees prior to data analyses}''. This cannot be true: $1/100$ degree of longitude is $921\,\mathrm{m}$ and
 that of latitude is $1{,}110\, \mathrm{m}$. Yet, the minimal distance between points in Figure 3 of \cite{MH16} is approximately $60\, \mathrm{m}$.
 This suggests that they have not performed ``pseudo-anonymization'' as claimed.
 
 \begin{figure}
    \begin{center}
      \includegraphics[width=4cm]{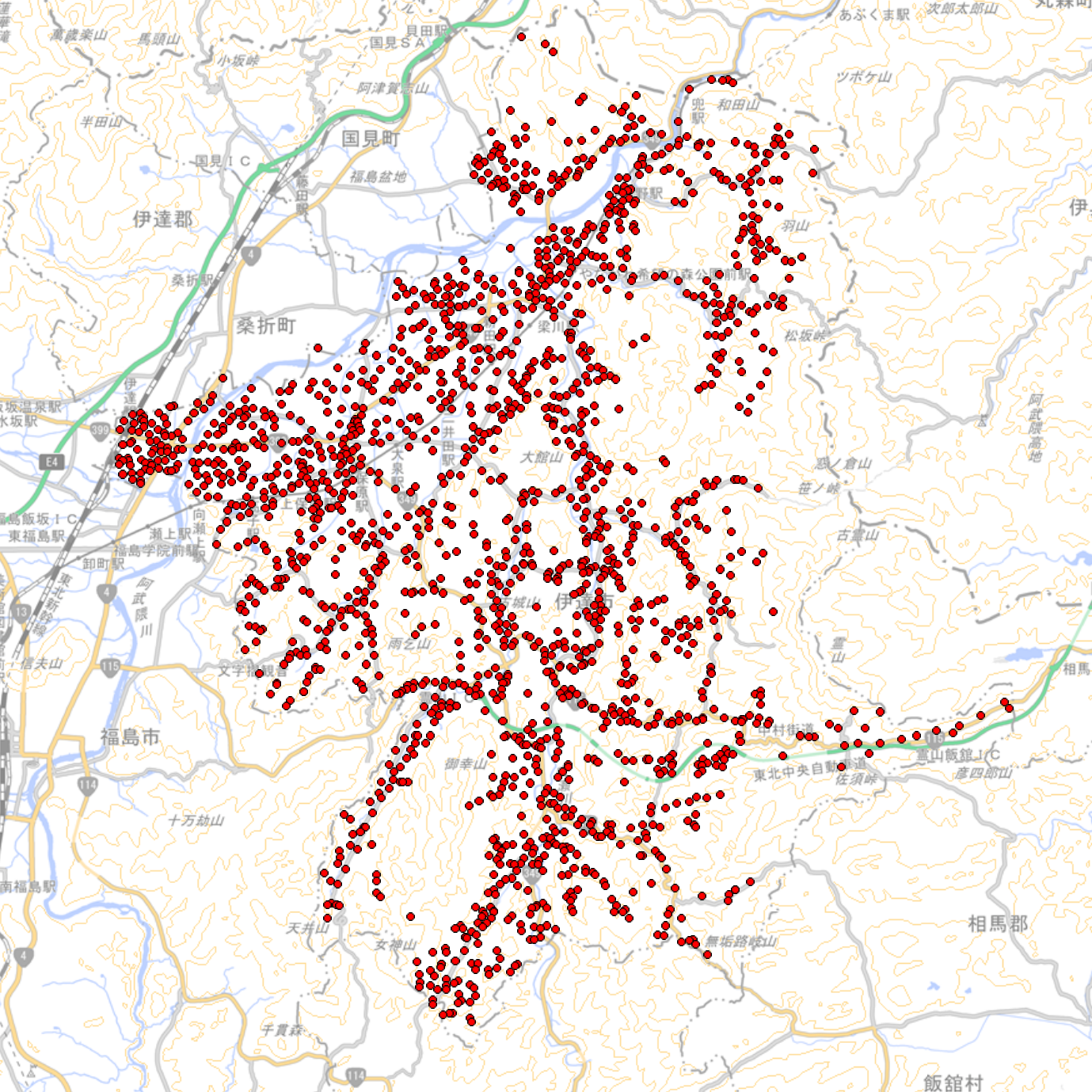}
      \includegraphics[width=4cm]{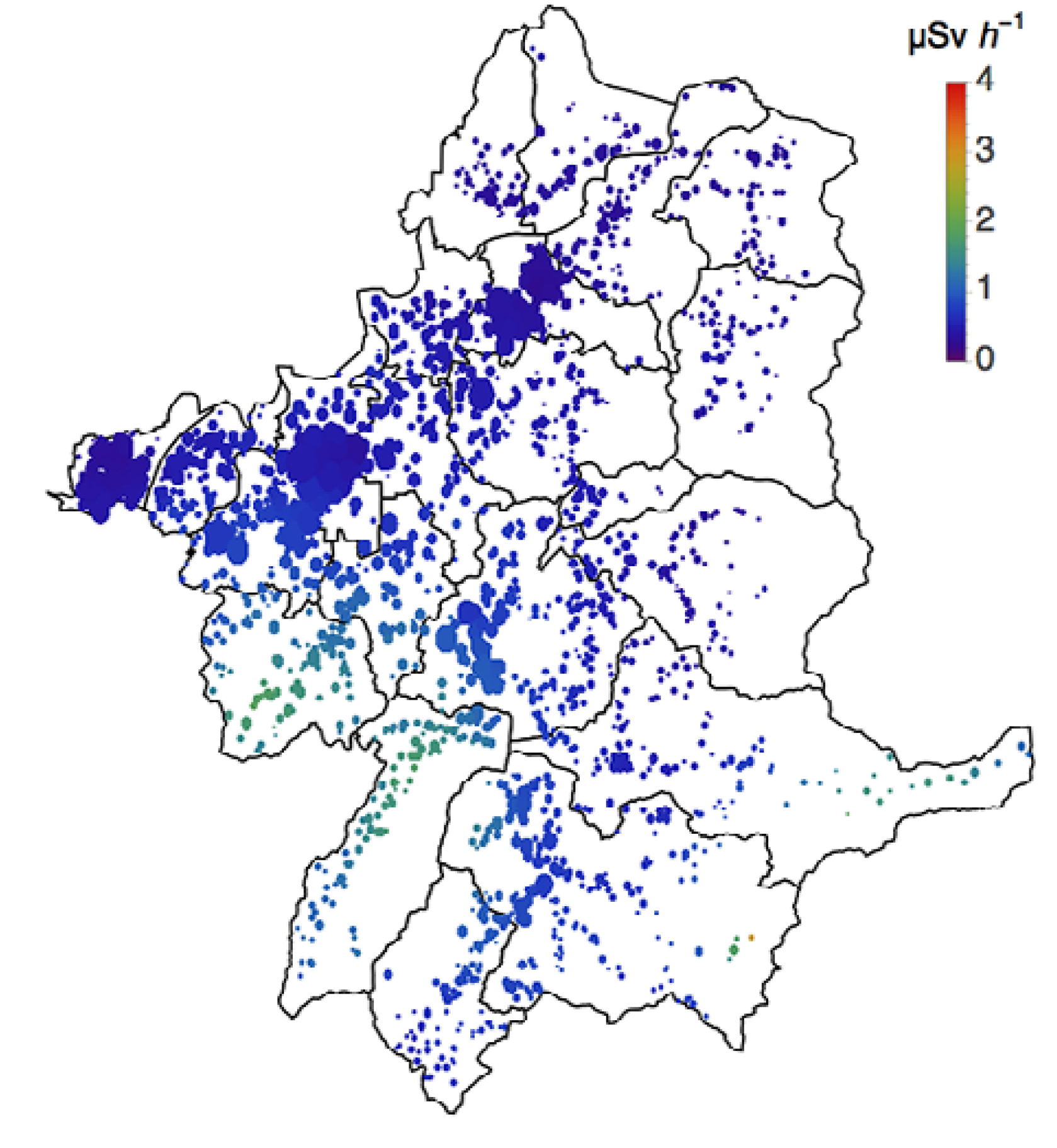}
      \includegraphics[width=4cm]{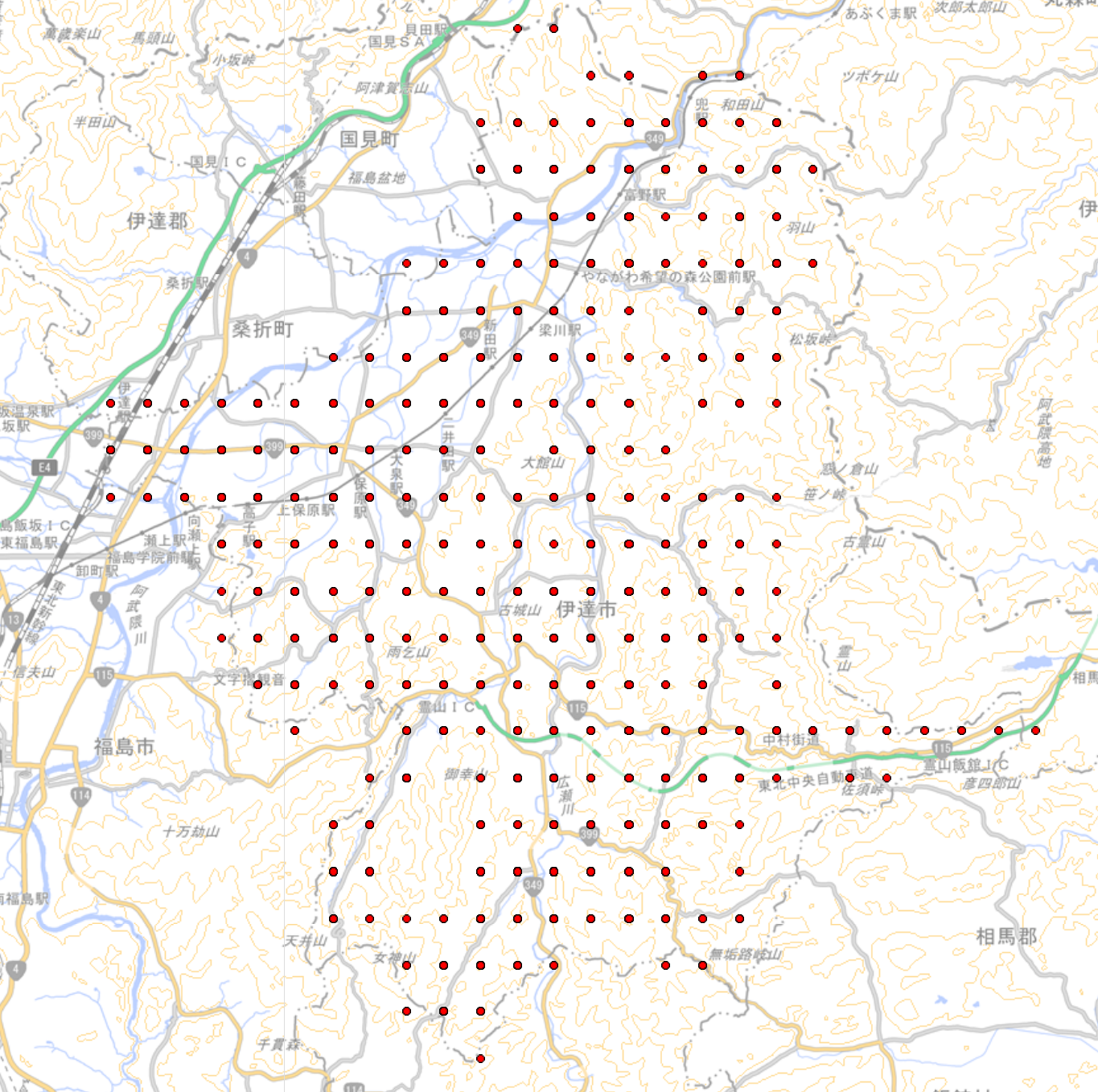}\\
     \includegraphics[width=4cm]{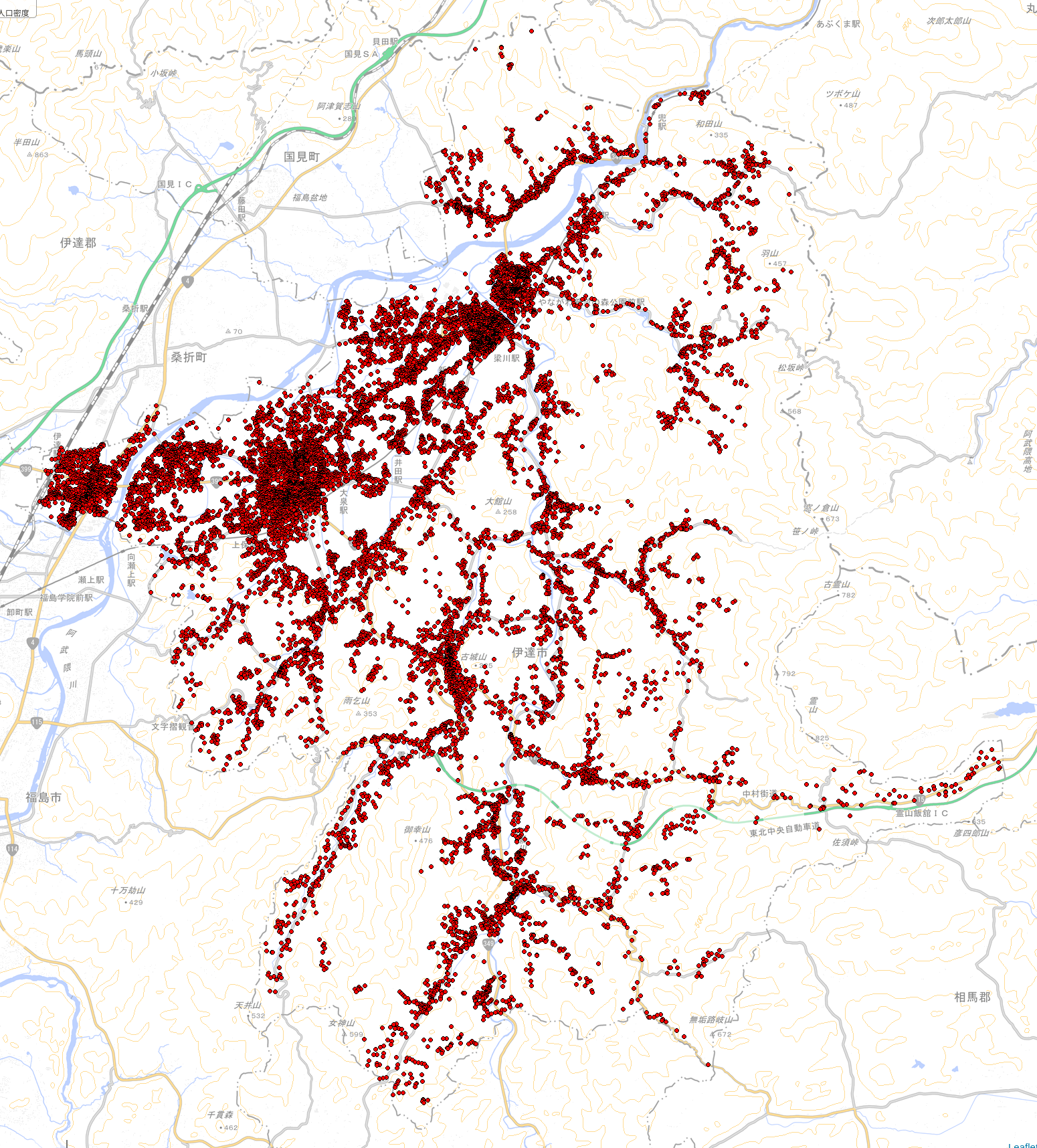}
     \includegraphics[width=4cm]{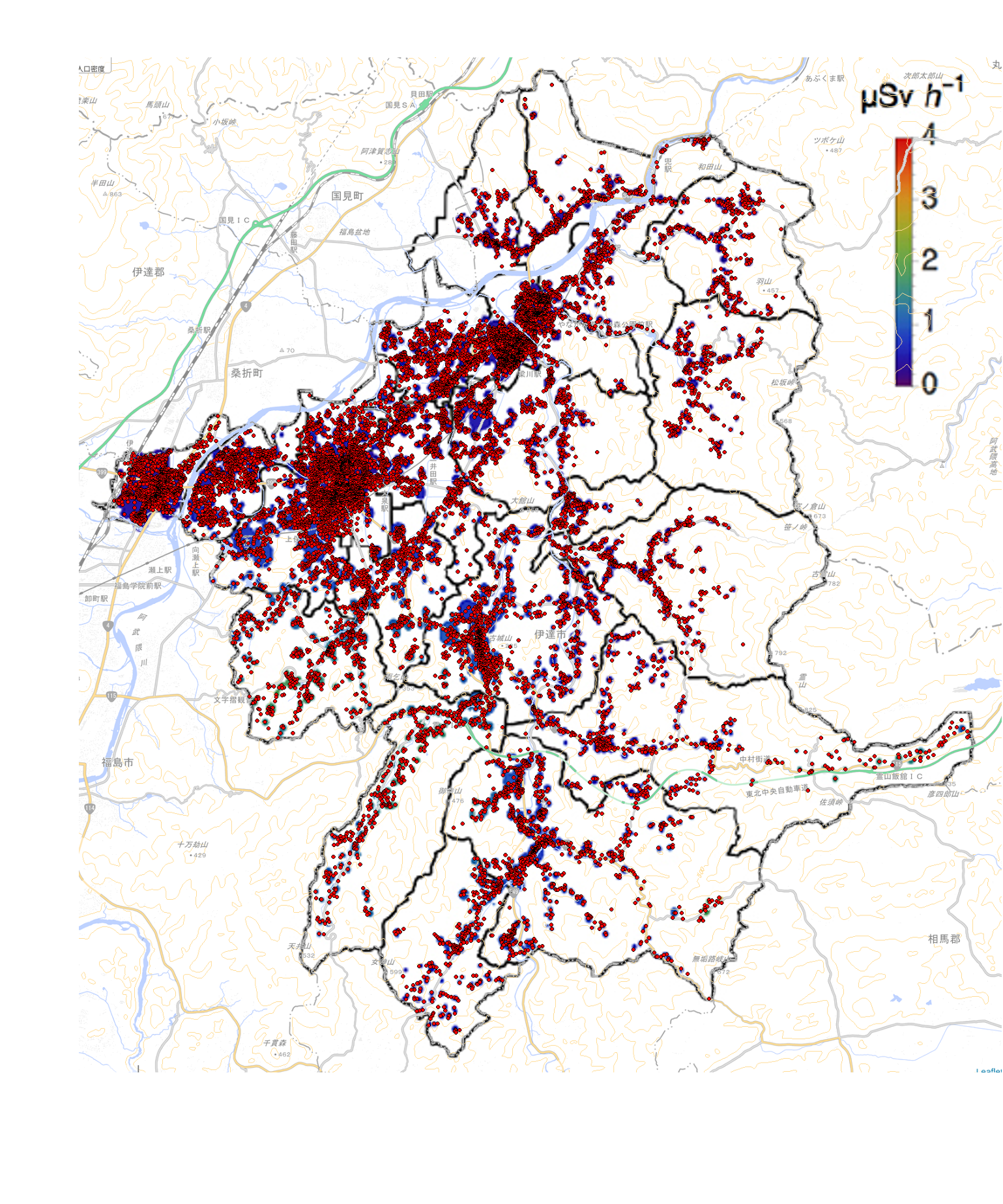}
    \end{center}
    \caption{  Plot of coordinate data from GIS and the plot from \cite{MH16}. Top left: GIS data of $1{,}800$ addresses (without street numbers),
    top center: Fig.\,3c of the paper \cite{MH16},
    top right:  same plot as top left but rounded to 1/100 degree as described in the paper, bottom left: GIS data with full street addresses
    and bottom right: GIS data with full street addresses superimposed on Fig.\,3c.
}
  \label{fig:map}
\end{figure}
Figure \ref{fig:map} shows the plot of the GIS data
down to the level
without the street number, one panel of
Figure 3 of \cite{MH16}, the plot of the GIS data rounded  in the
way described in the paper, and the plot of full street address.
It is clear that the plot of the paper does not look like a pseudo-anonymized
plot, but matches quite well with the original GIS data.
We made the GIS data using the public site
\url{http://ktgis.net/gcode/lonlatmapping.html}.

 \item The authors have computed the coefficient $c$ for different sets of population and it varied with age
 \cite{Date15July30}, without reporting it in the paper. On the other hand, the age distribution of the participants differs considerably
 depending on the period as shown in Table 1, but the paper ignores this difference.
 It is inappropriate to calculate $c$ simply by averaging this data set.

 \item In Fig.\,4a, the bins with $1.35,$ and $1.45\,\mu\svh$ contain less than $200$ and $400$ participants respectively,
 yet there are $4$ and $6$ points indicating upper outliers
 above the upper whisker.
 This would be impossible if the outliers were defined as those above the $99$-th percentile
 (there would be only two and four outliers, respectively), as claimed in the paper.
 There are similar issues in other figures.
 
 In fact, we see that the ratio between the upper whisker and the median of the personal dose rate
 is between $1.5$ and $3$ in the most populated bins
 (from $0.35$ to $0.55\,\mu\svh$ in Fig.\,4c, together
 contain more than $35{,}000$ participants).
 In the log-normal distribution matching Fig.\,5,
 2 and 3 times the median coefficient corresponds to $90$-th and $98$-th percentile, respectively.
We suspect that the upper whisker in Fig.\! 4 actually corresponds to a value much smaller than 99\%, such as 90\%. 

 If this is the case, there should be many participants above the upper whisker not presented in the Fig.\,4.
 For example, the upper whisker of the bin $2.15\,\mu\svh$ in Fig.\,4b (containing around $200$ participants)
 goes beyond the upper limit of the figure, meaning that around $20$ participants with doses rates higher than
 $1.05\,\mu\svh$ are omitted. It is inappropriate to omit such participants without any good reason.
 
 \item In Figs.\,4, there are bins containing only outliers ($0.45\,\mu \svh$ in Fig.\,4b,
 $0.25\,\mu \svh$ in Fig.\,4d and $0.25\,\mu \svh$ in Fig.\,4e).
 This would mean that $90\%$ (and not $99\%$, as discussed above)
 of the participants in these bins received $0\,\mathrm{mSv}$ in this period.
 This is highly unlikely (indeed most of the participants in the
 adjacent bins received non-zero doses).
 To confirm this, we looked at the dose data obtained from Date City by a Freedom of Information request and found that there were only 810 participants who received $0\,\mathrm{mSv}$ in the whole period corresponding to Fig.\,4b \cite{Date13DoseData}.
 The bin of ambient dose $0.45\,\mu\svh$ alone contains around $1{,}700$ participants, meaning that there were at least $1{,}500$ participants who received $0\,\mathrm{mSv}$, which contradicts with the fact that there are only 810 participants who received $0\,\mathrm{mSv}$.
 This discrepancy suggests a serious mistake in the analysis which might  affect all results in this paper.
 
 \item In Fig.\,5,  the cumulative distribution reaches $1$ (at least above 0.9999) at $c=1$, which
 implies there was no individual with $c>1$.
 This is in a clear contradiction with Figs.\,4, which show some participants with $c$ larger than $1$.
 
 \item Below Eq.\,(2), the authors state ``\textit{50-percentile is $c=0.15$}''.
 However, careful examination of Fig.\,5 shows that the $50$-th percentile (cumulative distribution 0.5) actually corresponds to $c=0.16$.
 Namely, the median of $c$ in Fig.\, 5 is $0.16$.
 Moreover, Fig.\,5 shows that the distribution of $c$ matches the log-normal distribution very well.
 The average over the log-normal distribution is greater than the median (see also item~\ref{item.lognormal} below).
 This immediately contradicts with the main result, Eq.\,(1), which states that the average of $c$ is $0.15$.
 The contradiction persists whether we use the median value $0.16$ from Fig.\,5 or $0.15$ from the authors' statement below Eq.\,(2).
 
 \item The above two issues suggest that ``outliers" might have been omitted from Fig.\,5 and
 the ``cumulative distribution" is defined with respect to the data excluding the ``outliers".
 If this is the case, there are two problems.
 First, the cumulative distribution is generally defined by including outliers.
 Moreover, there is no explanation how the outliers are defined and how many data are excluded from Fig.\,5.

 \item Below Eq.\,(1), the authors state ``\textit{$\langle \rangle$ denotes the average over all data in the six periods
 excluding the outliers}". However, the ``outliers'' are not defined here.
 The only place where the ``outliers'' are defined is in the first paragraph of Section 3, as an explanation
 of Fig.\,4. There, the ``outliers'' are defined as below $1$-st and above $99$-th percentile.
 However, as we have already discussed above, it already contradicts with Fig.\,4.
 If we assume that the same definition of ``outliers" applies to Eq.\,(1), the contradiction pointed out above
 becomes even more serious.
 That is, the log-normal distribution (logarithm of the random variable obeying the normal distribution
 with the mean $\mu$ and the standard deviation $\sigma$) matching Fig.\,5 is characterized by $\mu = \log{0.16}$ and $\sigma=0.54$.
 Its average over the section between $1$-st and $99$-th percentiles ($ 0.0455 < c < 0.562$)
 gives $c=0.18$, significantly larger than the claimed average $0.15$.
 \label{item.lognormal}
 
 \item In Section 3, in the main result $\langle c \rangle = 0.15 \pm 0.03$ (Eq.\,(1)), it is not explained what the
 claimed confidence interval(?) ``$\pm 0.03$'' exactly means, and how it is derived.

 \item In Section 4, the authors refer to the limitation of the study, but then go on to state ``\textit{[H]owever, we believe the differences in actual dosimeter use patterns among the
participants do not greatly affect the present results, as discussed below}''. This is misleading, to say the least, for two
reasons. First,
there is no quantitative analysis on the effect of the dosimeter use patterns to their result.
Second, the two papers cited, i.e. Refs.~\cite{Nomurae009555} and \cite{Naito2015}, do not directly apply to the situation addressed in this paper. In addition, both of them have several technical issues, and their conclusions themselves are not without problems, as shown in Hamaoka~\cite{YH19}.
 
 \item In the Abstract, the authors state ``\textit{The result show that the individual doses were about 0.15 times the ambient doses,
 the coefficient of 0.15 being a factor of 4 smaller than the value employed by the Japanese government, \ldots}''.
 However, even if all the data analyses in the paper were correct, the quoted value $0.15$ is the average (or the median?) of the coefficient $c$,
 which should not be used for the purpose of radiological protection and thus cannot be compared directly to the value employed by the government.
 In particular, the median value of $c$ means that the half of the population will actually have greater values of $c$.
 In fact, the data in Figs.\,4 and 5 show that the coefficient $c$ has a wide variation among individuals.
 International Commission on Radiological Protection (ICRP) recommends that
 ``\textit{the representative person should be defined such that the probability is less than approximately 5\%
 that a person drawn at random from the population will receive a greater dose}'' (Paragraph B50 of Ref.~\cite{ICRP101a}.)
 Following this recommendation, one should rather use the 95-th percentile value of $c$. 
 Assuming the data analyses in the paper are correct,
 from the log-normal distribution in Fig.\,5, the 95-th percentile is estimated as $c = 0.39$, which is more than half of the
 value $0.6$ employed by the Japanese government. We note that the authors do mention that the 90-th and 99-th
 percentile values are $c=0.31$ and $c=0.56$ (below Eq.\,(2)).

 \item In the  Conclusion section, the authors write ``\textit{it is possible to predict the external exposure
 dose received by each individual based on the aircraft monitoring data}''.
 As we pointed out above, individual fluctuations of the coefficient are not taken into account in
 drawing such a conclusion. The conclusion of this paper cannot be a basis of radiological protection.

 \end{enumerate}
 
 To summarize all the issues (1)--(12) above, the treatment of data in the paper does not satisfy the minimum standard required for scientific papers. For instance, it is even unclear whether the main conclusion ``$c=0.15$'' represents the average or the median of the actual data, and the number of samples excluded as outliers is not specified. Given that the exposure dose is known to follow the log-normal distribution, there should be no need to exclude part of the samples as outliers if the logarithm is taken in the first place. As observed in Fig.\,4, it is highly likely that $c$ is underestimated because samples with high dose exposure are excluded. The authors of this paper analyzed the logarithm of exposure in other papers  \cite{Adachi_2015}\cite{Nomurae009555}. It is necessary to explain why this paper did not do so.
 For the purpose of radiological protection, the large variation of the individual doses is important, and it is inappropriate just to use the median or the average.
 Furthermore, $c$ should depend on the age and lifestyles of participants, but the authors did not take that into account. Moreover, although they state ``\textit{[T]hese results show that coarse-grained airborne data can be a useful estimator for predicting the individual doses of residents living in contaminated areas}'' in Section 4, they do not give any information that shows the goodness-of-fit.

We note that, while one of us (S.\,K.) has written a comment~\cite{Kurokawa18} concerning the subsequent paper~\cite{MH17},
this Letter addresses separate issues on a different paper~\cite{MH16}.

\ifLetterStyle
\bigskip
\noindent
Yours Sincerely,

\medskip

\noindent
\myauthors
\fi

\subsubsection*{Acknowledgements}
We thank Ms.\! Akemi Shima for providing us with
the public documents obtained through her Freedom of Information requests.

Some of the issues pointed out in this Letter were discussed in Refs.~\cite{Kurokawa19,KT19-1,KT19-2}, in the Japanese magazine \textit{KAGAKU}.
This Letter is composed as an original article in English,
with a permission of the publisher of \textit{KAGAKU} (Iwanami Shoten, Publishers).
We also thank the \textit{KAGAKU} Editorial Office for opportunities to discuss this work.

\subsubsection*{Conflicts of interest}
Y.T.'s employment is funded through Programma per giovani ricercatori, anno 2014
``Rita Levi Montalcini'' of the Italian Ministry of Education, University and Research (MIUR).

{\small
\bibliographystyle{unsrt}
\bibliography{radiation}
}
\end{document}